# Group theory approach to unification of gravity with internal symmetry gauge interactions: II. Relativity of charges and masses

S.E. Samokhvalov

serg_samokhval@ukr.net

**Abstract.** The variant of electrogravitational unification has been studied on base of principle of relativity of charges and masses.

## 1. Introduction

Unification of gravity with other interactions falls out from the general picture of the unification of interactions. Interactions related to internal symmetries are united with each other based on the gauge principle by unification of the corresponding gauge groups. In contrast, the unification of gravity with other interactions is implemented on the basis of the Kaluza's multidimensional approach, or on the basis of supersymmetry. The difficulty of using the traditional scheme of a gauge approach in the case of external (space-time) symmetry is the reason for this difference [1]. Using the technique of deformed infinite Lie groups, in [2] a single group-theoretic approach was developed to describe the gauge fields of both internal and external symmetries. This allowed interpretation of gravity as a gauge theory of the deformed group of diffeomorphisms $\Gamma_T$ of space-time, which, in turn, allowed the electrogravitational unification to be realized on the basis of the gauge principle by unifying the gauge groups of gravity $\Gamma_T$ and electrodynamics $U(1)^g$ into one group, which is their semi-direct product $\Gamma_{T \otimes U(1)} = \Gamma_T \times) U(1)^g$ [3]. There are many ways to form semi-direct products of such groups, and hence the ways to form of this kind unifications. In [3], the simplest of them was presented: canonical variant. It leads to the Einstein equation with the energy-momentum tensor of an electromagnetic field as one of the sources and the Maxwell equations in the Riemannian space.

In this paper, the physical principles of another variant of the electrogravitational unification are formulated, which in a sense is more symmetrical than the canonical variant. Characteristic features of this variant of electrogravity (as opposed to the canonical variant) are: fundamental nonlinearity of electrodynamics, in particular the limitation of the electric potentials difference by the value of $\phi \approx 1{,}74 \cdot 10^{24}$ units SGS; the fundamental "smearness" of electric charges; the existence of additional degrees of freedom of the electrogravitational field. This new electrogravity becomes canonical for weak electromagnetic fields whose potentials are much smaller than $\phi$.

## 2. Principle of relativity of charges and masses

Let $S$ be a closed, ideally conductive shell, which we will call the frame $S$.

In Maxwell's theory, the following principle of relativity is valid: **no measurements of electromagnetic characteristics performed within the frame S can detect the electric potential of the frame S relative to any other frame S'.** We will call this principle the Maxwell's principle of relativity (MPR).

Let the frame $S$ contain a particle with a charge $q$ and a mass $m$. We move it to the frame $S'$, which has an electric potential $\upsilon$ relative to the frame $S$. In this case, the charge and mass of the particle are transformed:

$$q' = q, \qquad m' = m + q\upsilon/c^2. \tag{1}$$

These formulas indicate that the MPR is valid only to the extent, in which we can neglect changing the mass of the particle $q\upsilon/c^2$. This changing can be detected in processes where inertial or gravitational properties of particles are significant. Accuracy of accounting these properties defines the limits of the applicability of MPR.

The theory of electrogravity, which we propose, is based on the principle of relativity of charges and masses (PRCM), which extends the MPR in the following way: **no measurements of both electromagnetic and gravitational characteristics performed within the frame S can detect the electric potential of the frame S relative to any other frame S'.** In addition, **we postulate the invariance for one and the same particle**

$$z^2 = q^2 - (mc^2/\phi)^2, \tag{2}$$

**regardless of the frame in which it is located.** Here $\phi = \sqrt{2\pi/\kappa}$ is the fundamental constant of the electric potential dimension determined by the Einstein's gravitational constant $\kappa = 8\pi\gamma/c^4$. The negative sign in formula (2) and the value of the constant $\phi$ are uniquely determined from the condition of correspondence of the developing theory to the canonical electrogravity in the case of weak electromagnetic fields.

The formulated postulates require the replacement of transformations (1) to the following transformations:

$$q' = (q + mc^2\upsilon/\phi^2)/\beta, \qquad m' = (m + q\upsilon/c^2)/\beta, \tag{3}$$

as well as a change in the interpretation of dashed and non-dashed variables in them (here $\beta = \sqrt{1 - \upsilon^2 / \phi^2}$ ). To coordinate with the PRCM it is necessary to introduce an observer connected with the frame S (observer S). If one and the same particle is moved to different frames, then observers of these frames, according to PRCM, will attribute to this particle the same values of charge and mass, regardless of the electric potentials difference between the frames. The transformations (3) give a change of the particle charge and mass from a viewpoint of the observer $S$ when the particle moves from the frame $S$ to the frame $S'$ having a potential $\upsilon$ relative to $S$.

Observed values of charge and mass cannot be infinite or imaginary, hence there is restriction $\upsilon < \phi$, which indicates the fundamental nonlinearity of electrodynamics in the proposed theory.

Transformations (1) lie at the heart of the canonical variant of the electrogravitational unification (CEG) [3]. The electrogravity based on transformations (3) will be called the theory of relativity of charges and masses (TRCM), since the electric charges and the rest masses of the particles in it are no longer absolute, but depend on the conditions of their observation.

Let's note the analogy between the TRCM and the Einstein's theory of relativity (TR). Relative velocity in TR is similar to potentials difference in TRCM, and the principal impossibility of detecting the etheric wind corresponds to the impossibility of detecting the "vacuum" potential postulated in the TRCM. TR deprived ether of its dynamic properties, so the need for it disappeared. TRCM also translates some of the dynamic properties of physical vacuum into kinematics. For example, transformation (3) allows us to talk about "kinematic" renormalization of charges and masses.

For a relativistic generalization of transformations (3) let's write the invariant (2) using the energy-momentum vector $p_m$ of a particle:

$$z^2 = q^2 - (c / \phi)^2 p_m p^m .$$

The transformations that conserve this invariant form a group $SO(1,4) =: \Lambda^5$, which in addition to the transformations (3) also includes a four-dimensional Lorentz group $\Lambda^4$. The five quantities $r_a = (c p_m / \phi, q)$ ( $a$ takes the values 0, 1, 2, 3, 4, 5) behave like a covariant vector under the $\Lambda^5$ transforms. This vector is the source of the electrogravitational field in the TRCM, which is based on the $\Lambda^5$ invariance.

## 3. Theory of relativity of charges and masses

We formulate the TRCM as a gauge theory, using the method of deformed infinite Lie groups [2,3]. We note that the application of the traditional gauge approach is not possible in our case, since, firstly, the quantities related to external symmetries ($p_m$) are the sources of the gauge field, and secondly, the electric charges in the TRCM are not absolute, therefore, the geometric structure corresponding to the TRCM is not a structure of fibre bundle with connection.

Conservation $r_a$ is associated with $G = T \otimes U(1)$ symmetry ($T$ is a four-dimensional Abelian translation group), therefore, TRCM as a gauge theory is being constructed based on the Lie group $G$. In this case, in accordance with the general scheme of the deformations method, constructing is performed in the next sequence.

1. We construct generalized gauge group $\widetilde{\Gamma}_G = \widetilde{\Gamma}_T \rtimes) U(1)^g$ by introducing of dependence of elements of the group $G$ on points $x$ of the space-time $M$ and defining of special product law [2].

2. From the group of deformations of group $\widetilde{\Gamma}_G$, we distinguish the subgroup $D$ of deformations that do not violate the Riemann structure of a 5-dimensional manifold $P = M \times U(1)$ with a fundamental length $2\pi l$ along a compact fifth coordinate. Coefficients of such deformations

$$H_\alpha^a = \begin{pmatrix} h_\mu^m & -lV^m \\ -lkA_\mu & lv \end{pmatrix} \tag{4}$$

have a geometric meaning of the components of the coordinate vierbein relative to a pseudo-orthonormal vierbein on $P$, and $v = \sqrt{1 + V_m V^m}$, $lk = 1/\phi$, $k = e/c\hbar$. Here and thereafter Latin indices are vierbein (group) indices, and the corresponding Greek indices are coordinate ones.

Coefficients of deformations (4) are potentials of the gauge fields in the TRCM. Here additional (in comparison with the canonical electrogravity [3]) fields $V^m$ appeared and they are necessary to ensure the $\Lambda^5$ invariance of the theory.

3. The group $\Gamma_G$ is obtained from $\widetilde{\Gamma}_G$ with help of deformations belonging to $D$. The infinitesimal action of the group $\Gamma_G$ on $P$ is given by the formulas

$$\delta x^\mu = (h_m^\mu t^m + lkV^\mu A_m / w) t^m - lkcV^\mu \theta / w, \qquad \delta\varphi = -\frac{k}{w}(c\theta - A_m t^m), \tag{5}$$

in which $t^m$, $\theta$ are functions that parametrize the group $\Gamma_G$, $\varphi \in [0, 2\pi)$ is the fifth coordinate on $P$, and $w := v - lk A_m V^m$. Instead of $\theta$ it is sometimes convenient to use the parameter $t^5 = -lkc\theta$, and instead of $\varphi$ - the coordinate $x^5 = l\varphi$.

The generators $X_m$ and $X_\theta$ of the group $\Gamma_G$ action (5), as well as the operators of the energy-momentum $X_m$ and the charge $Q$, by which they are determined, have the following form:

$$X_m = h_m^\mu \partial_\mu - \frac{1}{c} A_m X_\theta, \qquad P_m = -i\hbar X_m,$$

$$X_\theta = -\frac{kc}{w}(\partial_\varphi + IV^\mu \partial_\mu), \qquad Q = -i\hbar X_\theta.$$

Attention is drawn to the fact that in the presence of an variable field $V^m$, the operators $P_m$ and $Q$ do not commute, that is, at the birth of the field $V^m$, the electric charge is smeared, but its total value remains unchanged due to the $\Gamma_G$ invariance of the theory that we postulate.

4. The strength of the gauge field in the gauge theory of group $\Gamma_G$ is characterized by its structure functions:

$$F_{kl}^m = \Phi_{kl}^m - lkA_\mu \, \delta_{kl}^{\mu\nu} h_\nu^n F_{n5}^m, \qquad\qquad F_{kl}^5 = lk(\Phi_{kl} - A_\mu \delta_{kl}^{\mu\nu} h_\nu^n F_{n5}^5), \qquad (6)$$

$$F_{k5}^m = (V^m{}_{,k} + \Phi_{kl}^m V^l)/w, \qquad\qquad F_{k5}^5 = (-V_m V^m{}_{,k}/v + lk\Phi_{kl}V^l)/w, \qquad (6')$$

where $\Phi_{\mu\nu}^m := -\partial_\mu h_\nu^m + \partial_\nu h_\mu^m$, $\Phi_{\mu\nu} := \partial_\mu A_\nu - \partial_\nu A_\mu$, $\delta_{kl}^{\mu\nu} := h_k^\mu h_l^\nu - h_l^\mu h_k^\nu$.

5. The principle of $\Gamma_G$ relativity is that the potentials of the gauge fields are determined with the accuracy of internal automorphisms of the group $\Gamma_G$, which are interpreted as gauge transformations of potentials:

$$h_\mu'^m = h_\mu^m + (F_{nk}^m \, h_\mu^k - F_{n5}^m \, lkA_\mu)\, t^n - \partial_\mu t^m + F_{n5}^m \, h_\mu^n \, lkc\theta, \qquad (7)$$

$$A_\mu' = A_\mu + (F_{mn}^5 \, h_\mu^m / lk + F_{n5}^5 A_\mu) t^n - F_{n5}^5 \, h_\mu^m \, c\theta - c\,\partial_\mu\theta, \qquad (7')$$

$$V'^m = V^m + (F_{nk}^m V^k - vF_{n5}^m) t^n + F_{k5}^m V^k lkc\theta. \qquad (7'')$$

The structure functions of the group $\Gamma_G$ (6) are invariant under transformations (7).

6. The Lagrangian of the gauge field in TRCM is constructed from the structure functions of the group $\Gamma_G$, which guarantees the $\Gamma_G$ invariance of the theory. In addition, it is necessary to provide the $\Lambda^5$ invariance of the theory, which is the basis of the TRCM. There is a three-parameter family of bilinear on structure functions Lagrangians, which are invariant under global $\Lambda^5$ transformations. Among them there is one

$$L_{hA} = \frac{c^4}{64\pi\gamma} w(F_{mn}^k F_{mn}^k + 2F_{mn}^k F_{kn}^m - 4F_{mn}^n F_{mk}^k - F_{kl}^5 F_{kl}^5$$

$$-2F_{k5}^l F_{k5}^l - 2F_{k5}^l F_{l5}^k + 4F_{m5}^m F_{m5}^m - 8F_{mn}^n F_{m5}^5 + 4F_{kl}^5 F_{kl}^l), \qquad (8)$$

which has right Newtonian limit and acquires a total divergence term under the local Lorentz $\Lambda^{5g}$ transformations:

$$h_\mu'^m = h_\mu^m + \omega^m{}_n h_\mu^n + lk\,\omega^{m5}A_\mu, \qquad\qquad A_\mu' = A_\mu - \omega^5{}_n h_\mu^n / lk, \qquad (9)$$

$$V'^m = V^m + \omega^m{}_n V^n + v\,\omega^{m5}. \qquad (9')$$

So the $\Lambda^{5g}$ invariance of the electrogravitational field equations takes place in this case. We choose $L_{hA}$ (8) as the Lagrangian of the TRCM. At $V^m \to 0$, it passes to the Lagrangian $L_{hA}^c$ of canonical electrogravity [3].

7. At the last stage, the identities of the second Noether theorem are taken into account, which follow from the gauge symmetries $\Gamma_G$ and $\Lambda^{5g}$ and determine both the quantities conserved in the TRCM and the structure of the equations of motion.

Displacement tensors in TRCM are defined as follows:

$$B_n^{\mu\nu} := \partial_{\partial_\nu h_\mu^n} L_{hA} = \frac{1}{\kappa}(w(\Gamma_{\cdot\cdot n}^{\mu\nu} + \delta_{nk}^{\mu\nu} R^k) + \delta_{lk}^{\mu\nu} V^l(K_{\cdot n}^k + \delta_n^k R + lk(A_n R_\cdot^k - \delta_n^k A_s R_\cdot^s))),$$

$$B^{\mu\nu} := \partial_{\partial_\nu A_\mu} L_{hA} = \frac{1}{2\pi lk}(wF^{\mu\nu} + \delta_{lk}^{\mu\nu} V^l(K_k - R_k)),$$

$$W_n^\nu := \partial_{\partial_\nu W^n} L_{hA} = -\frac{1}{\kappa}(K_{\cdot n}^\nu + (K_\cdot^\nu - R^\nu)V_n / v + h_n^\nu R + lk(A_n R^\nu - h_n^\nu A_k R_\cdot^k)),$$

where

$$\Gamma_{lkn} := \frac{1}{2}(F_{nlk} + F_{lnk} + F_{kln}), \qquad \Gamma_{kn} := \frac{1}{2}(-F_{kn}^5 + F_{kn5} + F_{nk5}), \qquad F_{kl} = \frac{1}{2}(F_{kl}^5 + F_{kl5} - F_{lk5}),$$

$$K_{kn} = \Gamma_{kn} + lkA_l \Gamma_{lkn}, \qquad K_k = F_{k5}^5 + lkA_l F_{lk}, \qquad R_k = F_{kl}^l + F_{k5}^5, \qquad R = F_{5m}^m.$$

With their help, we write the energy-momentum tensor $t_n^\nu$ and the vector of the electric current $\varepsilon^\nu$ of the electrogravitational field in the TRCM, which follow from the $\Gamma_G$ invariance of the theory and which are the Noether currents of the Lagrangian (8):

$$t_n^\nu = N_n^\nu - \frac{1}{c} A_n \varepsilon^\nu, \qquad \varepsilon^\nu = -lkc(N_n^\nu V^n + M_n^{\mu\nu} V_{,\mu}^n) / w, \qquad (10)$$

where

$$N_n^\nu = B_m^{\mu\nu} \Phi_{\mu n}^m - B^{\mu\nu} \Phi_{\mu n} + W_m^\nu V_{,n}^m - h_n^\nu L_{hA}, \qquad M_n^{\mu\nu} = B_n^{\mu\nu} + B^{\mu\nu} V_n / lk\nu.$$

The last of the formulas (10) indicates that, in the presence of the field $V^m$, the electric charge density in the TRCM can be different from zero even in the absence of charged particles.

As a consequence of the $\Gamma_G$ invariance of TRCM, there are strong Noether identities:

$$B_n^{\mu\nu} = -B_n^{\nu\mu}, \qquad\qquad B^{\mu\nu} = -B^{\nu\mu}, \qquad\qquad (11)$$

$$T_n^\nu = -\frac{1}{\sqrt{-g}} \partial_{h_\nu^n}(\sqrt{-g} L), \qquad J^\nu = -c \partial_{A_\nu} L, \qquad (12)$$

where $L = L_{hA} + L_\psi$ is the complete Lagrangian of the system of the electrogravitational field and the matter fields $\psi$, which interact with it, $T_n^\nu = t_n^\nu + \tau_n^\nu$, $J^\nu = \varepsilon^\nu + j^\nu$, $\tau_n^\nu = -\frac{1}{\sqrt{-g}} \partial_{h_\nu^n}(\sqrt{-g} L_\psi)$, $j^V = -c\partial_{A_V} L_\psi$. In addition, conservation laws are fulfilled on the extremals (weak Noether identities):

$$\partial_V(\sqrt{-g}\,T_n^V) = 0, \qquad \partial_V(\sqrt{-g}\,J^V) = 0. \tag{13}$$

All this determines the structure of the equations of the electrogravitational field in the TRCM, which is similar to the structure of the Maxwell equations:

$$\nabla_V B_m^{\mu V} = -T_m^\mu, \qquad \nabla_V B^{\mu V} = -\frac{1}{c} J^\mu, \qquad \nabla_V W_n^V = -S_n, \tag{14}$$

де where $S_n = \partial_{V^n} L$.

As a result of the $\Lambda^{5g}$ invariance of the TRCM, strong identities are fulfilled

$$\nabla_V(B_{[nm]}^{\cdot V} + W_{[n}^V V_{m]}) = 0, \qquad \nabla_V(lkA_\mu B_n^{\mu V} + B_n^{\cdot V}/lk + vW_n^V) = 0 \tag{15}$$

and on the extremals $\partial_V(\sqrt{-g}\,m_{ab}^V) = 0$, where $m_{ab}^V$ is the Noether current associated with $\Lambda^{5g}$ symmetry.

The system of equations of electrogravity (15) contains 24 equations with respect to variables $h_\mu^m$, $A_\mu$, $V^m$. $\Gamma_G$ invariance reduces the number of independent equations (14) by 5, and $\Lambda^{5g}$ invariance - by 10, which corresponds to the will in the choice of $\Gamma_G$ and $\Lambda^{5g}$ gauge conditions.

As a result of $\Gamma_G$ invariance, when the equations (14) hold, there are identities:

$$\nabla_\mu \tau_m^\mu = f_m - \frac{1}{c} A_m \nabla_\mu j^\mu, \qquad \nabla_\mu j^\mu = -lkc(f_m V^m + \alpha_n^V V_{,V}^n)/w, \tag{16}$$

where $f_m = \tau_n^V \Phi_{Vm}^n - \frac{1}{c} j^V \Phi_{Vm} - s_n V_{,m}^n$, $s_n = \frac{lk}{v} A_V \tau_n^V + \frac{1}{lkcv} j_n$, $\alpha_n^V = \tau_n^V + \frac{1}{lkcv} j^V V_n$.

The first of the equations (16) is the law of energy-momentum change of matter in the TRCM (for a macroscopic matter it is the equation of motion), while the second gives the non-conservation of the charge of matter in the TRCM in the presence of the field $V^m$. The complete charge of the system is conserved as a result of (13), so part of it is carried by the electrogravitational field. Equations (16) can be used to analyze the possibilities of the experimental finding of the effects of TRCM.

Although the TRCM is being constructed in this paper as a gauge theory based on four-dimensional concepts, it can also be constructed according to the Kaluza scheme by reducing the five-dimensional gauge theory of gravity (with the gauge group $\Gamma_{T\otimes U(1)}$) using the weaker cylindrical conditions than in [3]. For the TRCM the five-dimensional approach is more natural than in the case of canonical electrogravity, since the fifth coordinate becomes similar to four other coordinates due to the $A^5$ invariance of the TRCM.